# Superconducting Magnet System Concept with a Mechanical Energy Transfer in the Magnetic Field

Vladimir Kashikhin

*Abstract*— **There is an interest to design superconducting magnet systems working in a persistent current mode. These systems continuously generate magnetic field with disconnected power source working like permanent magnet devices. In this paper proposed a magnet system concept based on the direct mechanical energy transfer in the magnetic field. Short circuited superconducting coils do not have current leads and power source. To pump the mechanical energy in the superconducting coil used a magnetizer which magnetically coupled with the coil. The mechanical removing the magnetizer from the magnet induces a persistent current in the superconducting coil which generates the magnetic field. Iron dominated magnet system concept was investigated using OPERA3d code which confirmed a visibility of proposed approach.**

## I. Introduction

**M**OST of superconducting magnet systems energized from an external power source. A known approach is to use superconducting magnets working in a persistent current mode like in MRI solenoids [1]. Several high temperature superconducting (HTS) magnets were recently investigated at Fermilab [2] – [4]. Magnets worked during short period of time in a current transformer mode using a primary copper coil to pump the energy in a secondary short-circuited HTS coil. It allowed decouple warm primary and superconducting secondary eliminating also superconducting current leads, quench detection, protection systems, and continuously working power supply. In this paper investigated a novel concept to pump the mechanical energy in the magnetic field of dipole magnets.

## II. Magnetic and Mechanical Energies in Magnets

In general, accelerator magnets energized from power sources. But most electrical generators transfer a mechanical energy of rotor rotation through the rotational magnetic field to an electrical energy induced in a stator. In advanced electrical machines also used in rotors superconducting coils or permanent magnets. The principle of mechanical energy transfer to the magnetic field could be also used in accelerator magnet systems. It is useful to analyze this process and compare with a conventional electromagnet.

The proposed dipole magnet system concept shown in Fig. 1.

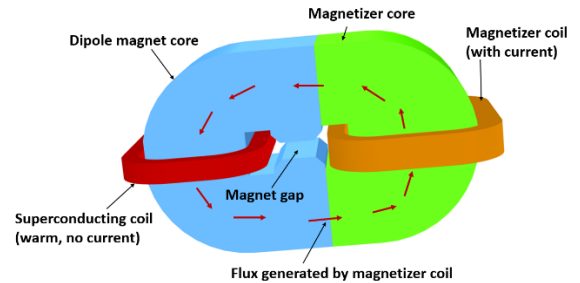

Fig. 1. Dipole magnet (left) with the powered magnetizer (right).

The C-type dipole magnet initially combined with another part named a magnetizer. The dipole magnet consists of the C-type ferromagnetic core with a short-circuited superconducting coil mounted on the magnet flux return yoke. To the gap side of magnet attached the magnetizer having also C-type core and a magnetization coil which could be non-superconducting.

The magnetizer initially generates the common magnetic flux with the dipole coil as shown in Fig. 1. At that time the dipole coil is in a non-superconducting state. After that the dipole coil cooled down to the superconducting state and the coil starts to operate in a "frozen flux" mode. The current in the magnetizer coil now reduced to zero and the magnetizer could be disconnected from the power source. This induces in an agreement with Lentz's law a continuously circulating current in the dipole superconducting coil, providing the magnetic flux constant condition (flux conservation law). In this case, the dipole coil total current equals the magnetizer previous total current. At this time the magnetizer should be mechanically removed from the dipole as shown in Fig. 2.

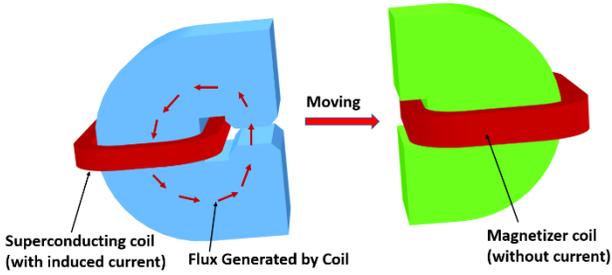

Fig. 2. Superconducting dipole magnet (left) with the unpowered removed magnetizer (right).

There are strong magnetic forces needed to separate both assemblies. But during this process all used mechanical energy transferred through the redirected magnetic flux in the dipole magnet to the dipole magnet stored energy concentrated in the magnet gap. The substantially increased dipole coil current continue to provide the magnetic flux constant condition. Most of mechanical energy used to the magnetizer removing transferred and concentrated in the magnet gap.

Because initially the ferromagnetic core forms the closed ferromagnetic loop without gaps there is needed a very low total current for the iron core magnetization. The proposed energy transfer process strongly depends on the material magnetic permeability. Fig. 3. shows a low carbon steel magnetic permeability and an analytic approximation for the field above 1 T.

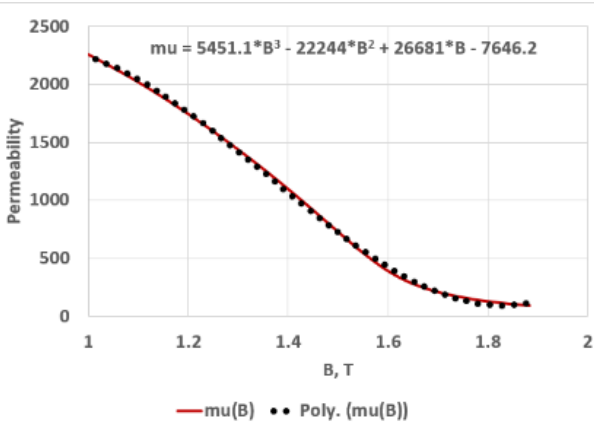

Fig. 3. Low carbon steel relative magnetic permeability and an approximation function.

From the simplified formulas we could understand the magnet parameters influence on the currents $I$, gap $\delta$, gap field $B\delta$, and the proposed approach limitations. The magnetizer total current $Icu$ for the closed magnetic circuit (see Fig. 1) and the superconducting dipole coil total current $Isc$ (see Fig. 2) could be defined:

$$Icu = \frac{Bfe \cdot Lfe}{\mu_o \mu}$$

$$Isc = \frac{Bfe \cdot Lfe}{2\mu_o \mu} + \frac{B\delta \cdot \delta}{\mu_o}$$

It is supposed that the magnetic flux is constant for both circuits and the $Lfe$ length of flux path for the closed circuit in the iron yoke is two times shorter than for the open one. The stored in the magnetic field energy transfer efficiency will be:

$$Kef = \frac{Wsc}{Wcu} = \frac{Isc}{Icu} = \frac{1}{2} + \frac{\mu(Bfe) \cdot B\delta \cdot \delta}{Bfe \cdot Lfe}$$

where $\mu(Bfe)$ is the iron magnetic permeability approximation shown in Fig. 3. At the fixed magnet geometry, the efficiency of the superconducting current increase is proportional the iron magnetic permeability and the magnet gap field (see Fig. 4).

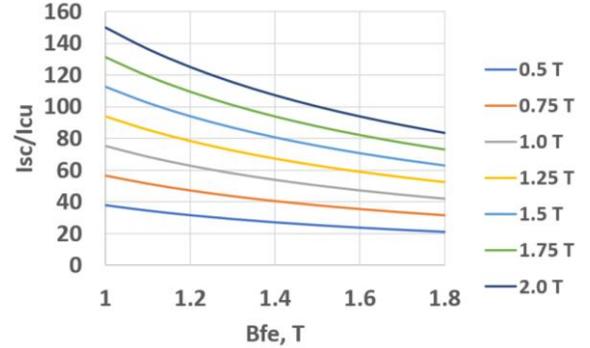

Fig. 4. Mechanical energy transfer efficiency for the different magnet gap (10 mm) fields.

Even for 1 T magnetic field in the gap which often used in accelerator magnets and 1.2 T in the iron the energy extracted from the magnetizer will be 60 times lower than the mechanical energy transferred in the magnetic field. If we increase the magnet gap two times the efficiency of energy transfer will increase also two times. Of course, the maximum energy transfer will be if all ferromagnetic material will be removed from the superconducting coil (air core magnet).

### III. MAGNETIZER

The magnetizer is a novel element used for accelerator magnets to pump the mechanical energy in the magnetic field. There are various types of magnetizers used in an industry named lifters to lift and transport ferromagnetic pieces. There are two general class of lifters: electromagnets [5] as shown in Fig. 2 (right) and permanent magnets [6] (see Fig. 5).

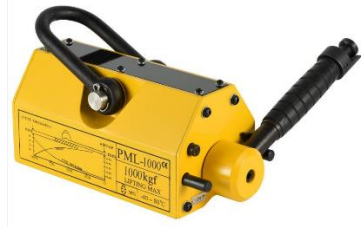

Fig. 5. Permanent magnet lifter for 1000 kg force.

This lifter has a handle to rotate the permanent magnet block inside the assembly to short-circuit the magnetic flux inside and eliminate the lifting force. Of course, the permanent magnet lifter magnetic circuit should be redesigned to eliminate the magnetic flux through the magnetizer provided by the superconducting coil. The most advanced lifters combines both approaches but for the separation used a short capacitor bank discharge in opposite direction to reduce the force needed for the separation.

Fig. 6 shows the dipole magnet with permanent magnet magnetizer.

One could see that the gap field follows the induced current in the superconducting loop. But the larger distance $dx$ the lower effect of separation. The peak force needed initially to separate the magnet and magnetizer is 330 kg (See Fig. 8) which exponentially decays with the separation distance. This value is in an agreement with the magnetic Maxwell pressure estimation for 1.16 T average field on separated surfaces:

$$Fx = \int_S \frac{B^2}{2\mu_0} \cdot d\mathbf{S}.$$

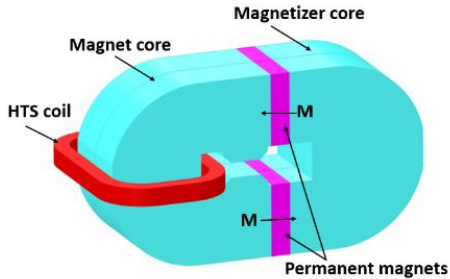

Fig. 6. HTS dipole magnet with a permanent magnet magnetizer. $M$ – vector of permanent magnet magnetization.

There could be various ways of magnetizer removing which discussed in the next section.

## IV. MECHANICAL ENERGY TRANSFER

The mechanical energy transfer process was investigated using OPERA3D software [5] and a short length (60 mm) dipole magnet model with the gap 10 mm. It is critical for this application to define the best way of mechanical energy transfer in the magnetic field by energizing the HTS coil. The well-known effect is that sliding magnetized objects relatively each other is much easier than pulling it out. In Fig. 7 shown the magnet model with the horizontal magnetizer moving (dx) as shown in Fig. 2 and in Fig. 7 – Fig. 8 the simulation results.

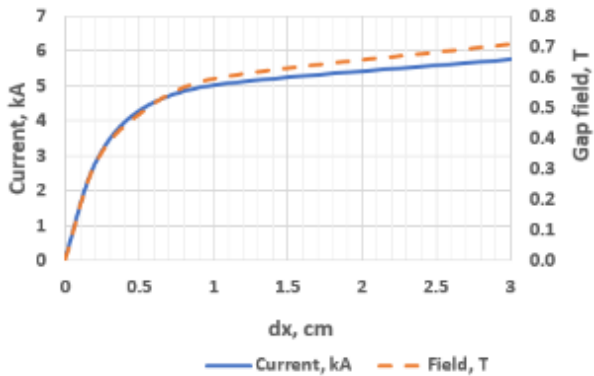

Fig. 7. Superconducting coil current and the gap field vs. the horizontal magnetizer displacement (dx).

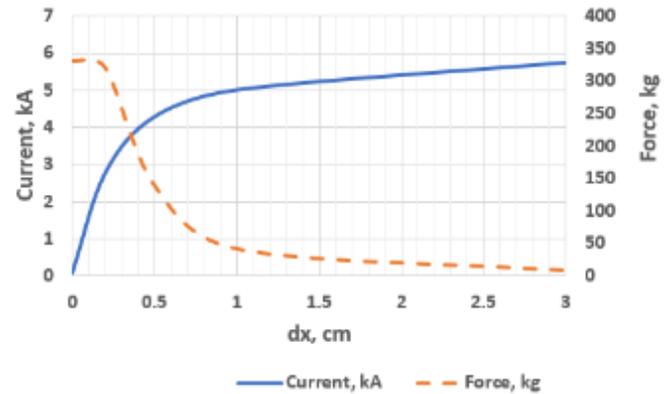

Fig. 8. Superconducting coil current and the magnetizer separation force vs. the horizontal magnetizer displacement (dx).

The other option for the separation is to move the magnetizer vertically as shown in Fig. 9. Before this moving the magnetizer coil current transferred in the superconducting coil and at this stage the magnetizer coil has a zero current, and this coil is not shown in Fig. 9.

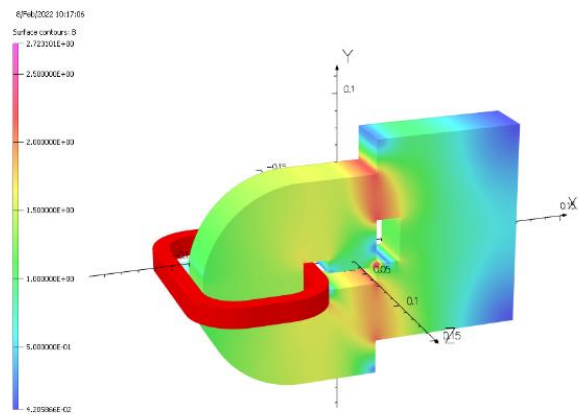

Fig. 9. Superconducting dipole magnet with the unpowered partially lifted magnetizer core.

The vertical magnetizer movement has slower current and gap field variation than the horizontal as shown in Fig. 10. But the total energy transfer is equal the energy transferred at the horizontal magnetizer movement.

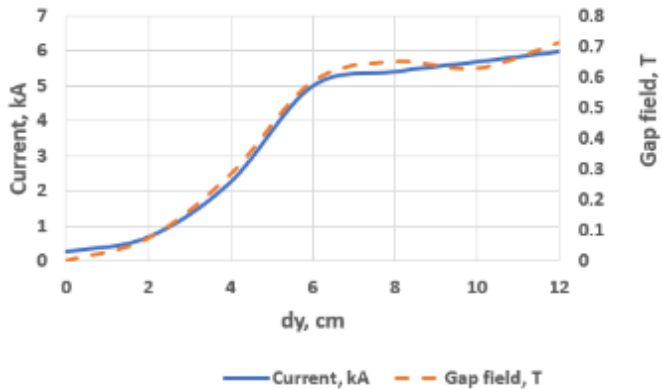

Fig. 10. Superconducting coil current and the gap field vs. the vertical magnetizer displacement (dy).

In this case the gap field and induced current have a lower variation with the separation distance. Fig. 11 shows the current and the separation force variation.

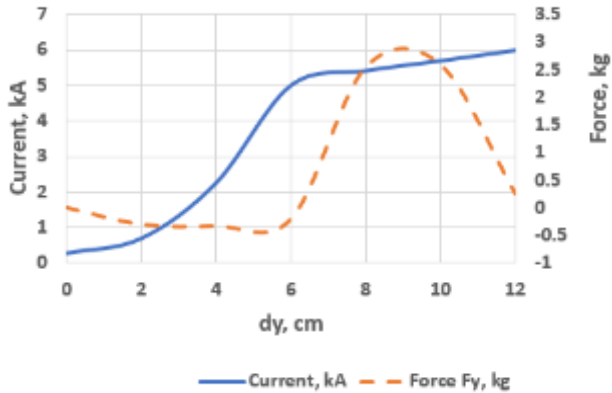

Fig. 11. Superconducting coil current and the separation force vs. the vertical magnetizer displacement (dy).

The main advantage of vertical magnetizer moving is that the needed peak separation force in hundred times lower than for the horizontal movement but at the same value of mechanical energy transfer.

## V. Comparison with a Conventional Electromagnet

Relatively low magnetic field HTS magnets could replace for some applications conventional room temperature electromagnets or permanent magnets. But reduced operational power losses in HTS magnets comes from the superconductivity by using cryogenics. It is not obvious which approach is better from the point of magnet system efficiency. The previously described HTS dipole magnet was redesigned as a room temperature magnet with a water-cooled copper coil. The magnet has also 10 mm gap and the peak field in the gap 0.7 T. For this field needed the coil with the total current 5.6 kA. The optimal current density for these types of magnets is 4 A/mm$^2$. In this case the copper cross-section will be 1400 mm$^2$. The coil power dissipation will be 0.9 kW for the 1 m magnet length. At the average cost of electricity in US is 0.12 $/kWh, the operational cost of this magnet will be 0.11 $/h or 964 $/year. The liquid nitrogen production cost is 0.18 $/litter. So, the HTS magnet cryostat evaporation rate could be 0.6 l/h to make even operational costs. Of course, it is a very rough estimation which is not included for the room temperature magnet costs of power supply, cabling, water cooling system, protection, and monitoring systems. But, at least, it makes a visible choice of HTS magnets using at low magnetic fields for some applications.

## Conclusion

The proposed and investigated HTS magnet concept opens a way to pump the mechanical energy in the magnetic field of accelerator dipole magnet. For that used a removable magnetizer which could be an electromagnet or a permanent magnet. It is critical for this type of magnet the way of magnet and magnetizer separation. The magnetic field and forces analysis showed that the separation across flux lines needs order of magnitude lower separation peak force than along flux lines on separated surfaces. Brief comparison with a conventional room temperature magnet makes visible the choice of HTS magnet systems for low field applications. The proposed mechanical energy transfer concept was successfully verified by building and testing HTS dipole magnet [8].

## References

[1] J. Minervini, M. Parizh, M. Schippers, " Recent Advances in Superconducting Magnets for MRI and Hadron Radiotherapy", Supercond. Sci. Technol. 31 (2018) 030301.
[2] V. Kashikhin, J. DiMarco, A. Makarov, Z. Mendelson, S. Rabbani, S. Solovyov, D. Turrioni, "High Temperature Superconducting Quadrupole Magnets with Circular Coils", *IEEE Trans. on Applied Superconductivity*, 2019, Vol. 29, Issue 5, 4002404.
[3] V. Kashikhin, D. Turrioni, "HTS Quadrupole Magnet for the Persistent Current Mode Operation", *IEEE Trans. on Applied Superconductivity*, VOL. 30, NO. 4, JUNE 2020,4602104.
[4] V. Kashikhin, D. Turrioni, "HTS Dipole Magnet Model for the Persistent Current Operation", *IEEE Trans. on Applied Superconductivity*, VOL. 32, Issue: 6, February 2022, 21643599.
[5] "Electromagnetic Lifters", SUMAN Enterprises, https://www.sumanenterprises.net/electromagnetic-lifters.html
[6] "Magnetic Lifters", Vestil, https://www.vestil.com/product.php?FID=598
[7] OPERA USER GUIDE, Dassault Systemes, UK Ltd, February 2020.
[8] V. Kashikhin, D. Turrioni, "HTS Dipole Magnet with a Mechanical Energy Transfer in the Magnetic Field", (published at this conference).